\documentclass[iop,apj]{emulateapj}
\usepackage{graphicx,url}

\shorttitle{Hydra II, Pisces II, and Laevens 1}
\shortauthors{Kirby, Simon, \& Cohen}
\slugcomment{Accepted to ApJ on 2015 July 28}
\begin{document}

\newcommand{\vmeanh}{ 303.1}
\newcommand{\vmeanerrh}{   1.4}
\newcommand{\sigmavh}{1.4}
\newcommand{\sigmaverrh}{1.3}
\newcommand{\sigmaverrlowerh}{1.0}
\newcommand{\sigmaverrupperh}{1.6}
\newcommand{\sigmavlimoneh}{3.6}
\newcommand{\sigmavlimtwoh}{4.5}
\newcommand{\sigmavlimthreeh}{6.7}
\newcommand{\sigmavmodoneh}{6.4}
\newcommand{\sigmaverrlowermodoneh}{1.8}
\newcommand{\sigmaverruppermodoneh}{2.4}
\newcommand{\sigmavlimmodtwoh}{6.0}
\newcommand{\sigmavmodthreeh}{6.7}
\newcommand{\sigmaverrlowermodthreeh}{1.9}
\newcommand{\sigmaverruppermodthreeh}{2.4}
\newcommand{\sigmavlimmodfourh}{3.6}
\newcommand{\sigmavlimmodfiveh}{3.6}
\newcommand{\vnmhone}{ 319.0}
\newcommand{\vnmhoneerr}{ 2.3}
\newcommand{\vnmhonediff}{15.9}
\newcommand{\vnmhonesigmamod}{2.0}
\newcommand{\vnmhtwo}{ 315.7}
\newcommand{\vnmhtwoerr}{ 3.7}
\newcommand{\vnmhtwosigma}{9.2}
\newcommand{\vnmhtwodiff}{12.6}
\newcommand{\sigmafehruleouth}{79}
\newcommand{\nh}{13}
\newcommand{\fehmeanwh}{-2.02}
\newcommand{\fehmeanwerrh}{0.08}
\newcommand{\fehmeanh}{-2.20}
\newcommand{\fehmeanerrh}{0.30}
\newcommand{\fehmeanerrlowerh}{0.35}
\newcommand{\fehmeanerrupperh}{0.25}
\newcommand{\fehsigmah}{0.40}
\newcommand{\fehsigmaerrh}{0.37}
\newcommand{\fehsigmaerrlowerh}{0.26}
\newcommand{\fehsigmaerrupperh}{0.48}
\newcommand{\sigmafehlimoneh}{1.10}
\newcommand{\sigmafehlimtwoh}{1.47}
\newcommand{\mlh}{ 30}
\newcommand{\mlerrlowerh}{ 20}
\newcommand{\mlerrupperh}{ 40}
\newcommand{\mllimoneh}{200}
\newcommand{\mllimtwoh}{320}
\newcommand{\vmeanp}{-226.5}
\newcommand{\vmeanerrp}{   2.7}
\newcommand{\sigmavp}{5.4}
\newcommand{\sigmaverrp}{3.0}
\newcommand{\sigmaverrlowerp}{2.4}
\newcommand{\sigmaverrupperp}{3.6}
\newcommand{\mlruleoutp}{99}
\newcommand{\sigmafehruleoutp}{85}
\newcommand{\np}{7}
\newcommand{\fehmeanwp}{-2.45}
\newcommand{\fehmeanwerrp}{0.07}
\newcommand{\fehmeanp}{-2.44}
\newcommand{\fehmeanerrp}{0.35}
\newcommand{\fehmeanerrlowerp}{0.30}
\newcommand{\fehmeanerrupperp}{0.40}
\newcommand{\fehsigmap}{0.48}
\newcommand{\fehsigmaerrp}{0.49}
\newcommand{\fehsigmaerrlowerp}{0.29}
\newcommand{\fehsigmaerrupperp}{0.70}
\newcommand{\sigmafehlimonep}{1.56}
\newcommand{\sigmafehlimtwop}{2.28}
\newcommand{\mlp}{370}
\newcommand{\mlerrlowerp}{240}
\newcommand{\mlerrupperp}{310}
\newcommand{\vmeanc}{ 149.3}
\newcommand{\vmeanerrc}{   1.2}
\newcommand{\sigmavc}{1.6}
\newcommand{\sigmaverrc}{1.4}
\newcommand{\sigmaverrlowerc}{1.1}
\newcommand{\sigmaverrupperc}{1.7}
\newcommand{\sigmavlimonec}{3.9}
\newcommand{\sigmavlimtwoc}{4.8}
\newcommand{\sigmavlimthreec}{7.0}
\newcommand{\sigmavmodc}{2.7}
\newcommand{\sigmaverrlowermodc}{1.4}
\newcommand{\sigmaverruppermodc}{1.8}
\newcommand{\vcrtblue}{ 155.3}
\newcommand{\vcrtblueerr}{ 1.8}
\newcommand{\vcrtbluesigma}{3.7}
\newcommand{\vcrtbluediff}{ 6.0}
\newcommand{\sigmafehruleoutc}{37}
\newcommand{\nc}{10}
\newcommand{\fehmeanwc}{-1.68}
\newcommand{\fehmeanwerrc}{0.05}
\newcommand{\fehmeanc}{-1.72}
\newcommand{\fehmeanerrc}{0.12}
\newcommand{\fehmeanerrlowerc}{0.15}
\newcommand{\fehmeanerrupperc}{0.10}
\newcommand{\fehsigmac}{0.14}
\newcommand{\fehsigmaerrc}{0.14}
\newcommand{\fehsigmaerrlowerc}{0.11}
\newcommand{\fehsigmaerrupperc}{0.18}
\newcommand{\sigmafehlimonec}{0.40}
\newcommand{\sigmafehlimtwoc}{0.53}
\newcommand{\mlc}{ 20}
\newcommand{\mlerrlowerc}{ 20}
\newcommand{\mlerrupperc}{ 20}
\newcommand{\mllimonec}{130}
\newcommand{\mllimtwoc}{190}
\newcommand{\vgsrh}{ 135.4}
\newcommand{\vgsrp}{ -79.9}
\newcommand{\vgsrc}{   4.6}

\newcommand{\vsyserr}{1.49}
\newcommand{\gcmeanv}{-144.4}
\newcommand{\gcmeanverr}{1.3}
\newcommand{\gcsigmav}{7.4}
\newcommand{\gcsigmaverrlower}{0.9}
\newcommand{\gcsigmaverrupper}{1.0}
\newcommand{\ngc}{52}
\newcommand{\ntemplate}{47}

\newcommand{\lvdiff}{2.6}
\newcommand{\lvdifferr}{2.2}
\newcommand{\lonefeh}{-1.55}
\newcommand{\lonefeherr}{ 0.11}
\newcommand{\ltwofeh}{-1.65}
\newcommand{\ltwofeherr}{ 0.11}

\title{Spectroscopic Confirmation of the Dwarf Galaxies Hydra~II and Pisces~II and the Globular Cluster Laevens~1}

\author{Evan~N.~Kirby\altaffilmark{1},
  Joshua~D.~Simon\altaffilmark{2},
  Judith~G.~Cohen\altaffilmark{1}}

\altaffiltext{*}{The data presented herein were obtained at the
  W.~M.~Keck Observatory, which is operated as a scientific
  partnership among the California Institute of Technology, the
  University of California and the National Aeronautics and Space
  Administration. The Observatory was made possible by the generous
  financial support of the W.~M.~Keck Foundation.}
\altaffiltext{1}{California Institute of Technology, 1200 E.\ California Blvd., MC 249-17, Pasadena, CA 91125, USA}
\altaffiltext{2}{Observatories of the Carnegie Institution of Washington, 813 Santa Barbara Street, Pasadena, CA 91101, USA}


%%%%%%%%%%%%%%%%%%%%%%%%%%%%%%%%%
%%%%%%%%%    ABSTRACT    %%%%%%%%
%%%%%%%%%%%%%%%%%%%%%%%%%%%%%%%%%

\begin{abstract}

We present Keck/DEIMOS spectroscopy of stars in the recently
discovered Milky Way satellites Hydra~II, Pisces~II, and Laevens~1\@.
We measured a velocity dispersion of
$\sigmavp_{-\sigmaverrlowerp}^{+\sigmaverrupperp}$~km~s$^{-1}$ for
Pisces~II, but we did not resolve the velocity dispersions of Hydra~II
or Laevens~1\@.  We marginally resolved the metallicity dispersions of
Hydra~II and Pisces~II but not Laevens~1\@.  Furthermore, Hydra~II and
Pisces~II obey the luminosity--metallicity relation for Milky Way
dwarf galaxies ($\langle {\rm [Fe/H]} \rangle = \fehmeanwh \pm
\fehmeanwerrh$ and $\fehmeanwp \pm \fehmeanwerrp$, respectively),
whereas Laevens~1 does not ($\langle {\rm [Fe/H]} \rangle = \fehmeanwc
\pm \fehmeanwerrc$).  The kinematic and chemical properties suggest
that Hydra~II and Pisces~II are dwarf galaxies, and Laevens~1 is a
globular cluster.  We determined that two of the previously observed
blue stars near the center of Laevens~1 are not members of the
cluster.  A third blue star has ambiguous membership.  Hydra~II has a
radial velocity $\langle v_{\rm helio} \rangle = \vmeanh \pm
\vmeanerrh$~km~s$^{-1}$, similar to the leading arm of the Magellanic
stream.  The mass-to-light ratio for Pisces~II is
$\mlp_{-\mlerrlowerp}^{+\mlerrupperp}~M_{\sun}/L_{\sun}$.  It is not
among the most dark matter-dominated dwarf galaxies, but it is still
worthy of inclusion in the search for gamma rays from dark matter
self-annihilation.

\end{abstract}

%%%%%%%%%%%%%%%%%%%%%%%%%%%%%%%%%
%%%%%%%%%   SECTION 1   %%%%%%%%%
%%%%%%%%%%%%%%%%%%%%%%%%%%%%%%%%%

\section{Introduction}
\label{sec:intro}

Ultra-faint dwarf galaxies (UFDs) harbor a wealth of information about
dark matter and nucleosynthesis.  They are the most dark
matter-dominated objects known \citep{sim07}.  Their large
dark-to-luminous mass ratios and their small sizes imply large central
densities of dark matter.  Hence, they are excellent targets for the
detection of gamma rays from dark matter self-annihilation
\citep{bon15}.  They also contain the largest concentrations of
metal-poor stars of any galaxy type \citep{kir08b,fre14}.  The most
metal-poor stars in UFDs could be the direct descendants of the first
generation of stars in the Universe.  If so, then their compositions
are direct samples of Population~III nucleosynthesis \citep{fre10b}.

The Sloan Digital Sky Survey \citep[SDSS,][]{aba09} revolutionized the
field of dwarf galaxies a decade ago by discovering the first UFDs.
In total, more than a dozen new UFDs were found in the SDSS
\citep[e.g.,][]{wil05,zuc06,bel07}.  In the span of roughly five
years, SDSS and its successor, the Sloan Extension for Galactic
Understanding and Evolution \citep[SEGUE,][]{yan09}, more than doubled
the number of known Milky Way satellites.  The field is experiencing
another rejuvenation with the arrival of several new imaging surveys
with deeper photometry and/or coverage of the Southern sky: the Dark
Energy Survey (DES), the Survey of the MAgellanic Stellar History
\citep[SMASH,][]{nid15}, the Panoramic Survey Telescope \& Rapid
Response System \citep[Pan-STARRS,][]{kai10}, ATLAS at the VLT Survey
Telescope \citep{sha15}, and independent imaging with the Dark Energy
Camera (DECam) at the CTIO/Blanco telescope.  To date, DECam imaging,
including DES, has enabled the discovery of 12 new Milky Way
satellites \citep{kop15a,bec15,kim15a,kim15b,kim15c,kim15d}.
Pan-STARRS discovered two satellites \citep{lae14,lae15}, and ATLAS
and SMASH have each discovered one
\citep{bel14,mar15}\footnote{\citet{lae14}, using Pan-STARRS,
  discovered Laevens~1 simultaneously with \citet{bel14}, who used
  ATLAS\@.}.  Together, these surveys discovered 15 additional
satellites in less than a year and a half.  Their success can be
attributed to deeper photometry and surveying a different part of the
sky than SDSS\@.

The photometric discovery is the first step in identifying a new dwarf
galaxy.  In order to classify a stellar system as a galaxy, it should
exhibit some evidence for dark matter in the form of a large
mass-to-light ratio, metallicity dispersion, or both \citep{wil12}.
These criteria can be tested only with spectroscopy.  Two of the newly
discovered satellites, Reticulum~II and Horologium~I, have been
spectroscopically confirmed as dwarf galaxies by both velocity and
metallicity dispersions \citep{sim15,wal15,kop15b}.  The other systems
have tentative classifications as galaxies or globular clusters (GCs)
based on their luminosities and half-light radii ($r_h$).  GCs have
$r_h < 30$~pc regardless of luminosity \citep{har96}, but the $r_h$ of
galaxies increases with luminosity \citep[e.g.,][]{bel07}.  The
half-light radii of dwarf galaxies overlap with GCs ($r_h \approx
30$~pc) at $M_V \approx -2$, and they grow to $r_h > 100$~pc at $M_V <
-5$.

This study concerns three recently discovered satellites.
\citet{mar15} discovered Hydra~II in DECam images taken as part of
SMASH\@.  Its large half-light radius, $68 \pm 11$~pc, strongly
suggests that it is a dwarf galaxy.  Hydra~II is especially
interesting for its proximity to the Magellanic Clouds.
\citeauthor{mar15}\ raised the possibility that it is associated with
the Large Magellanic Cloud (LMC), which is potentially true for many
of the DES-discovered satellites \citep{kop15a,bec15,dea15}.
\citet{bel10} discovered Pisces~II in SDSS images, and they confirmed
an overdensity of main sequence turn-off stars with deeper KPNO/MOSAIC
images.  Its size, $r_h \approx 60$~pc, suggests that it is also a
galaxy.  However, no spectroscopy has been obtained since its
discovery.  \citet{bel14} and \citet{lae14} co-discovered
Laevens~1/Crater\footnote{We call the object Laevens~1, as is the
  convention for GCs.}  with ATLAS and Pan-STARRS, respectively.  Its
half-light radius is only about 20~pc, and its luminosity is $M_V
\approx -5$.  Therefore, it can be tentatively classified as a GC, but
spectroscopy is required for a definitive identification.

We observed Hydra~II, Pisces~II, and Laevens~1 with Keck/DEIMOS in
order to identify them as galaxies or clusters.  We completed the
identification by measuring both velocity and metallicity dispersions.
We also quantified the metallicities of a few stars in these galaxies
to get a hint of their capability to enhance their own metallicities.

%%%%%%%%%%%%%%%%%%%%%%%%%%%%%%%%%
%%%%%%%%%   SECTION 2   %%%%%%%%%
%%%%%%%%%%%%%%%%%%%%%%%%%%%%%%%%%

\section{Photometry and Astrometry}
\label{sec:phot}

In order to design DEIMOS slitmasks, we needed a catalog of
coordinates for stars that were potentially members of the three
satellites.  We also needed colors and magnitudes of stars to identify
candidate members.  \citet{san12} published a photometric and
astrometric catalog for Pisces~II, but no such published catalog
exists for Hydra~II or Laevens~1\@.  We downloaded publicly available
images for Hydra~II, and we obtained new Keck/LRIS images of Laevens~1\@.

\subsection{Hydra~II}

\citet{mar15} discovered Hydra~II in SMASH/DECam images.  The Hydra~II
images, taken on 2013 March 20, are publicly available through the
NOAO Science Archive\footnote{\url{http://www.portal-nvo.noao.edu/}}.
We downloaded the calibrated images in the DECam $g$ and $r$ filters.
These images are flat fielded, and they have astrometry headers.  We
used SExtractor \citep{ber96} to identify stars and measure their
magnitudes in each of the $g$ and $r$ images.  We discarded objects
with $\texttt{class\_star} \le 0.8$ in order to weed out galaxies and
image artifacts.  We corrected magnitudes for extinction according to
the dust maps of \citet[][SFD98]{sch98}.

\begin{figure*}[t!]
\centering
\includegraphics[width=0.7\textwidth]{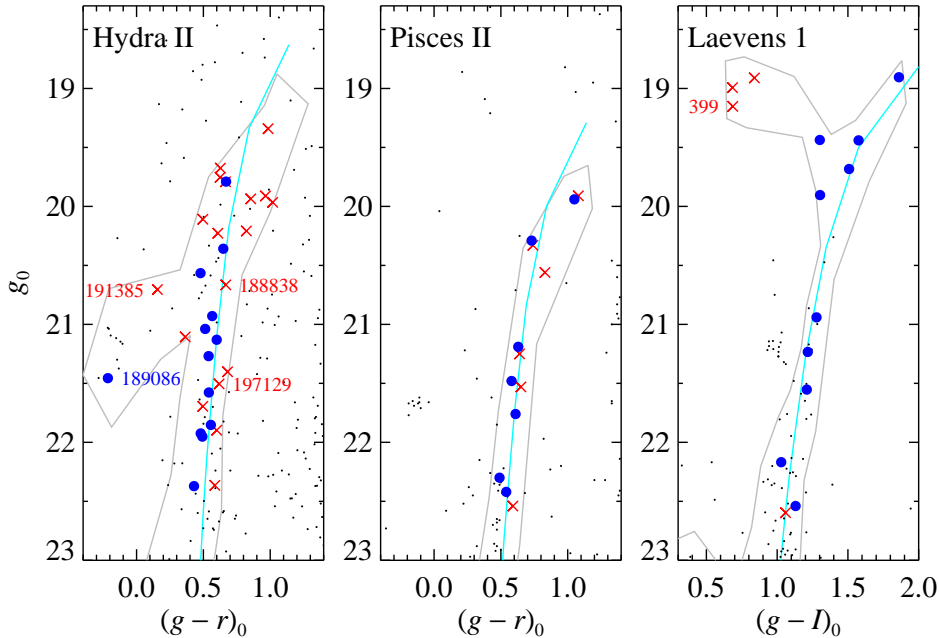}
\caption{Color-magnitude diagrams of the three satellites.  Blue
  points are spectroscopically confirmed members, whereas red crosses
  are non-members.  Small black points are stars within $4'$ of the
  satellite's center that we did not observe spectroscopically.  The
  gray regions circumscribe the stars that were considered possible
  members.  The ridgeline for the metal-poor GC M92 \citep{cle06} is
  shown in cyan.\label{fig:cmds}}
\end{figure*}

\begin{figure*}[t!]
\centering
\includegraphics[width=0.9\textwidth]{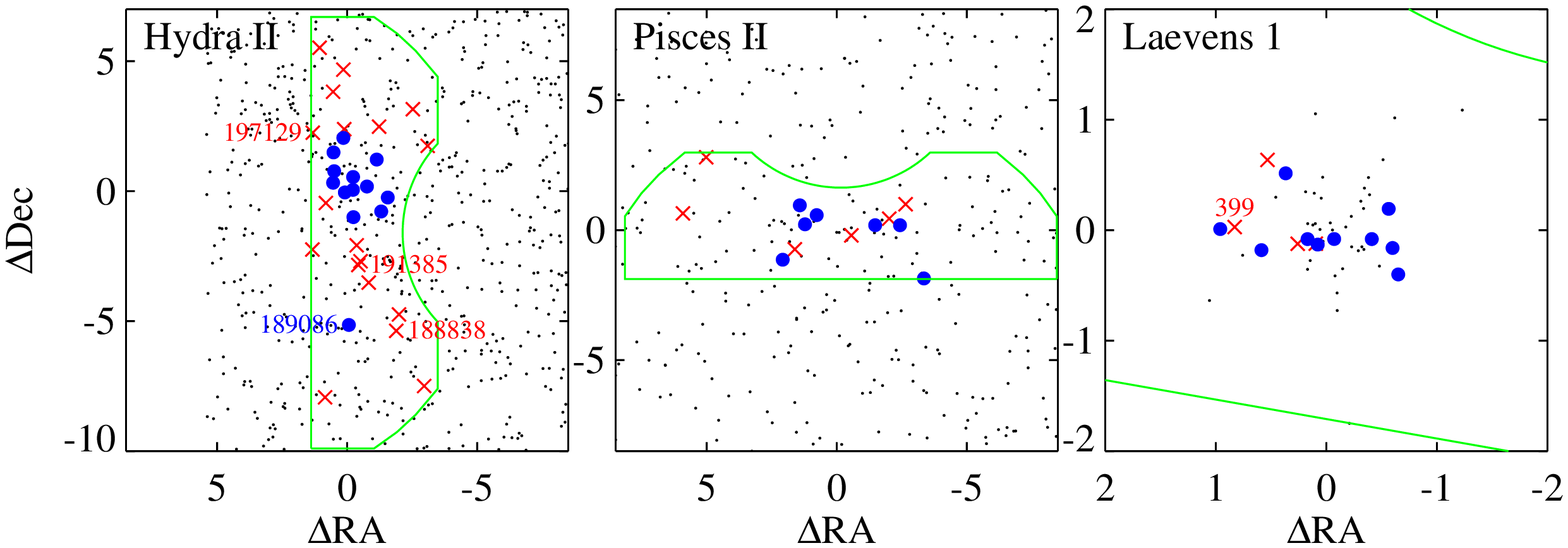}
\caption{The sky coordinates of spectroscopic targets.  The values
  shown are displacements from the satellite center, as measured by
  \citet{mar15}, \citet{bel10}, and \citet{lae14}.  Blue points are
  spectroscopically confirmed members, whereas red crosses are
  non-members.  Small black points are stars with $g_0 < 22$ not
  targeted for spectroscopy.  The DEIMOS slitmask outline is shown in
  green.  Because the field for Laevens~1 is smaller than the other two
  satellites, the axis ranges are smaller.\label{fig:maps}}
\end{figure*}

We selected stars for spectroscopy based on their colors and
magnitudes.  We drew a region in the color--magnitude diagram (CMD) in
the approximate shape of the red giant branch (RGB) and horizontal
branch (HB)\@.  Figure~\ref{fig:cmds} shows this region in gray.  We
assigned priorities for spectroscopic selection within the region
based on magnitude but not on color.  These priorities were used to
resolve conflicts where slitmask design constraints forced a choice
among two or more stars.  Brighter stars were given higher priorities.
The stars that were able to be placed on the DEIMOS slitmask are shown
as blue points (members) and red crosses (non-members).
Figure~\ref{fig:maps} shows the sky coordinates of the spectroscopic
targets.

Our spectroscopic selection includes two potential HB stars, 189086
and 191385, and one asymptotic giant branch (AGB) star, 194563\@.  Our
kinematic membership selection (Section~\ref{sec:membership}) includes
all three stars.  However, star 191385 is 0.4~mag above the HB\@.
Therefore, we ruled it a non-member.  Star 189086 has a color and
magnitude consistent with the HB.  Although it is the most distant
member in our Hydra~II sample, we kept it in our list of members.
These decisions do not affect our results in a measurable way
(Section~\ref{sec:sigmav}).

\subsection{Pisces~II}

\citet{san12} observed Pisces~II with Magellan/Megacam in the $g$ and
$r$ filters.  We used their published astrometric and photometric
catalog.  They corrected for extinction using the SFD98 dust map.  We
drew a spectroscopic selection region around the RGB
(Figure~\ref{fig:cmds}).  Because Pisces~II has a smaller half-light
radius \citep[$1\farcm 1$,][]{san12} than Hydra~II \citep[$1\farcm
  7$,][]{mar15}, the possible targets are denser on the sky, which
causes more conflicts for spectroscopic selection.  As a result, the
selection region did not include the HB so that RGB stars could be
selected instead.  (RGB spectra lend themselves more easily to the
measurements of radial velocity and metallicity than HB spectra.)  As
for Hydra~II, spectroscopic priority was given to brighter stars.

\subsection{Laevens~1}

D.~Perley kindly observed Laevens~1 for us with Keck/LRIS
\citep{oke95} on 2015 March 23.  Simultaneous exposures were obtained
for 130~s in the $g$ filter in the blue arm and 120~s in the $I$
filter in the red arm.  The images were reduced with
\texttt{LPipe}\footnote{\url{http://www.astro.caltech.edu/$\sim$dperley/programs/lpipe.html}},
which provides flat fielding, astrometric solutions, and photometric
calibration.

We identified stars and measured their magnitudes with DAOPHOT
\citep{ste87,ste11}.  The LRIS point spread function (PSF) was
asymmetric, and it varied over the field.  We allowed DAOPHOT to
choose the best analytic PSF, which was a Penny function (a
two-dimensional Gaussian core with Lorentzian wings).  Additionally,
there was a look-up table that allowed the PSF to vary linearly over
the field.  Again, we corrected for extinction using the SFD98 dust
map.

Because the half-light radius of Laevens~1 \citep[$0\farcm
  5$--$0\farcm 6$,][]{bel14,lae14} is even smaller than Pisces~II, we
again selected stars on the RGB, not the HB\@.  However, we also
included three bright, blue stars near the center of the system.  The
nature of these stars is controversial.  \citet{bel14} speculated that
these are core helium-burning blue loop stars, which would imply that
Laevens~1 has formed stars within the last few hundred Myr, making it
a dwarf galaxy.  \citet{lae14} also identified these stars, but they
did not favor their interpretation as blue loop stars because young,
blue stars at fainter magnitudes were absent.  We obtained spectra of
all three stars in order to determine if they are members of
Laevens~1\@.  We also note that \citet{bonif15} found a population of
stars brighter and bluer than the main sequence turn-off but too faint
for spectroscopy.  If they are not blue stragglers, these stars could
signify the presence of a $\sim$2~Gyr old population in Laevens~1.

For the purposes of measuring metallicities, it is convenient to have
photometry in a uniform system.  Therefore, we converted Cousins $I$
magnitudes into SDSS $i$ magnitudes following the conversion formula
of \citet{jor06}: $i = I + 0.21(R-I) + 0.34$.  The formula depends
weakly on $R-I$ color, which we approximated as 0.5~mag for all RGB
stars.  Due to the imprecision of this assumption, we added 0.05 in
quadrature to the $i$ photometric errors.  This error floor
corresponds to an error in $R-I$ color of 0.2~mag.  For comparison,
the full range of $R-I$ color for an old, metal-poor RGB is about
0.4~mag.

%%%%%%%%%%%%%%%%%%%%%%%%%%%%%%%%%
%%%%%%%%%   SECTION 3   %%%%%%%%%
%%%%%%%%%%%%%%%%%%%%%%%%%%%%%%%%%

\section{Spectroscopy}
\label{sec:obs}

\subsection{Observations}

\begin{deluxetable*}{lcccclc}
\tablewidth{0pt}
\tablecolumns{7}
\tablecaption{DEIMOS Observations\label{tab:obs}}
\tablehead{\colhead{System} & \colhead{UT Date} & \colhead{\# targets} & \colhead{Airmass} & \colhead{Seeing} & \colhead{Individual Exposures} & \colhead{Total Exposure Time} \\
\colhead{ } & \colhead{ } & \colhead{ } & \colhead{ } & \colhead{($''$)} & \colhead{(s)} & \colhead{(s)}}
\startdata
\cutinhead{Milky Way Satellites}
M22          & 2009 Oct 13 & 64 & 1.5  & $0.6$ & $2 \times 900$ & 1800 \\
             & 2009 Oct 14 & 64 & 1.5  & $0.6$ & $2 \times 900 + 420$ & 2220 \\
Hydra~II     & 2015 May 18 & 49 & 1.6  & $0.8$ & $4 \times 1200 + 1 \times 900$ & 5700 \\
Pisces~II    & 2015 May 18 & 20 & 1.8  & $0.8$ & $3 \times 1380$ & 4140 \\
Laevens~1    & 2015 May 18 & 14 & 1.4  & $0.9$ & $3 \times 1380$ & 4140 \\
Laevens~1 (star 1717) & 2015 Mar 22 & \phn 1 & 1.2  & $0.6$ & $1 \times 1800$ & 1800 \\
\cutinhead{Radial Velocity Standard Stars}
HD~38230      & 2012 Apr 19 & \nodata & 1.4  & $1.0$ & $1 \times 120$ & \phn 120 \\
HD~151288     & 2013 Apr 13 & \nodata & 1.1  & $1.0$ & $1 \times 80$ & \phn \phn 80 \\
HD~103095     & 2013 May 5\phn & \nodata & 1.1  & $0.7$ & $1 \times 60$ & \phn \phn 60 \\
BD$-18^{\circ}$~5550 & 2013 May 18 & \nodata & 1.3  & $0.8$ & $1 \times 350$ & \phn 350 \\
HD~88609      & 2015 May 18 & \nodata & 1.2  & $0.8$ & $1 \times 240$ & \phn 240 \\
HD~122563     & 2015 May 18 & \nodata & 1.4  & $0.8$ & $1 \times 75$ & \phn \phn 75 \\
HD~187111     & 2015 May 18 & \nodata & 1.2  & $0.9$ & $1 \times 80$ & \phn \phn 80 \\
BD$+23^{\circ}$~3912   & 2015 May 18 & \nodata & 1.0  & $0.9$ & $1 \times 300$ & \phn 300 \\
HD~109995     & 2015 May 19 & \nodata & 1.1  & $0.9$ & $1 \times 300$ & \phn 300 \\
\cutinhead{Telluric Standard Stars}
HR~4829       & 2015 May 19 & \nodata & 1.1  & $0.9$ & $1 \times 60$ & \phn \phn 60 \\
HR~7346       & 2015 May 19 & \nodata & 1.1  & $0.7$ & $1 \times 120$ & \phn 120 \\
\enddata
\end{deluxetable*}

We observed one slitmask for each satellite with DEIMOS \citep{fab03}
on 2015 May 18.  Table~\ref{tab:obs} lists the exposure times of each
slitmask.  We used the 1200G grating with a ruling of 1200
lines~mm$^{-1}$ and a blaze wavelength of 7760~\AA\@.  The grating was
tilted such that the center of the CCD mosaic corresponded to
7800~\AA\@.  Slits were $0.7\arcsec$ wide.  This configuration gives
an approximate wavelength range of 6300--9100~\AA\ at a resolution of
1.3~\AA\ FWHM ($R \sim 6000$ at 7800~\AA)\@.  The exact wavelength
range of each spectrum depends on the location of the slit on the
slitmask.  The starting and ending wavelengths of the spectra vary by
$\sim 300$~\AA\ across the slitmask.  We also obtained internal flat
field and arc lamp exposures in the afternoon and morning for calibration.

As mentioned above, \citet{bel14} identified several bright, blue
stars in the vicinity of Laevens~1\@.  These stars would be unusual in
an old GC or dwarf galaxy.  Our slitmask included two of these stars.
We observed another blue star, 1717, with a single $0.7''$ slit.

We also observed nine radial velocity standard stars and two telluric
(hot) stars with a long slit.  The slit width was $0.7''$.  We used a
long slit that spanned the entire length of the DEIMOS field of view.
This allowed us to refine the wavelength solution based on night sky
lines over the entire CCD mosaic.  Section~\ref{sec:reductions}
describes our approach to the wavelength solution.

\subsection{Reductions}
\label{sec:reductions}

As in our previous papers \citep[e.g.,][]{sim07,sim11,kir13a,kir15},
we reduced the DEIMOS spectra using a slightly modified version of the
\texttt{spec2d} IDL data reduction pipeline developed by the DEEP2
team \citep{coo12,new13}.  We introduced a few updates to those
procedures for this data set.  The updates include improvements to the
wavelength solution as determined from sky emission lines and
corrections for the effects of differential atmospheric refraction.
In the first stage of the sky line wavelength tweaking, a zero point
offset is determined during the main reduction pipeline.  We improved
the tracing of the sky lines across each slit at low S/N, which is
particularly important for our long-slit observations of bright
template stars with short integration times.  We also added a second
stage to the sky line fitting: after the reductions are completed, a
quadratic fit is performed to the sky line wavelengths in each
extracted spectrum, and then the variation of each of the quadratic
fit parameters as a function of position on the slit mask is fit with
a polynomial.  This process corrects for errors in the flexure
compensation system as well as temperature changes between the times
at which the arc frames and the science frames were obtained (Geha et
al., in prep.)\@.

Atmospheric refraction shifts the position of the star light within
the slit during the observations.  Shifts in the spatial direction
(along the slit) result in curvature of the object profile on the
detector, and we changed the DEEP2 extraction algorithm to account for
this effect \citep{kir15}.  Shifts in the wavelength direction (across
the slit) result in velocity offsets that vary as a function of
wavelength.  Using the airmass at the time of observation, the angle
between the slit and the parallactic angle, and the measured seeing,
we computed a correction for this velocity shift and applied it to the
extracted spectra.

In order to take maximum advantage of these improvements, we
constructed a new empirical library of template stars for radial
velocity measurements to replace the template set most DEIMOS dwarf
galaxy studies have employed since \citet{sim07}.  We selected a
sample of metal-poor stars spanning a range of effective temperature,
surface gravity, and metallicity, and we observed them by orienting
the slit north-south and slowly driving the telescope such that the
star moved steadily across the slit.  This process ensured that the
star light uniformly illuminated the slit during the exposure.  To
make sure that the template star observations contained strong enough
sky lines for accurate adjustments to the wavelength solution (as
described above), we integrated for at least 60~s during each template
observation even if the star crossed the slit in a much shorter amount
of time.  We also replaced the telluric template from \citet{sim07}
with higher S/N observations using the same techniques.

\begin{figure}[t!]
\centering
\includegraphics[width=\columnwidth]{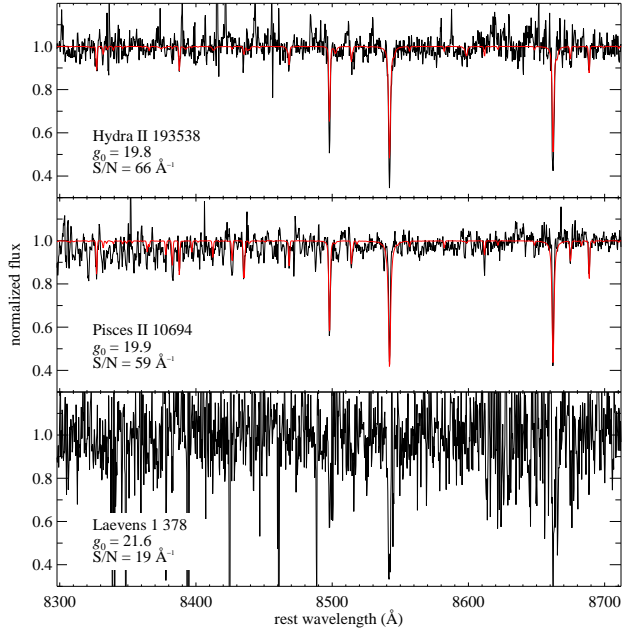}
\caption{Example spectra for one red giant in each of the satellites.
  The top two spectra have high S/N, and the bottom spectrum has low
  S/N\@.  We fit model spectra (red) to the high-S/N spectra
  (Section~\ref{sec:feh}).  Pisces~II 10694 is carbon-rich with CN
  absorption visible between 8300~\AA\ and 8400~\AA\@.  The model
  spectrum has a normal carbon abundance, so there is excess CN
  absorption in the observed spectrum compared to the
  model.\label{fig:spectra}}
\end{figure}

Figure~\ref{fig:spectra} shows example 1-D spectra of one member star
in each satellite.  The spectra shown for Hydra~II and Pisces~II have
high S/N, while the spectrum for Laevens~1 has low S/N\@.  Also shown
are the best-fit model spectra (described in Section~\ref{sec:feh})
for the stars with S/N high enough to measure metallicity.  Because
the S/N is low for the star in Laevens~1, we did not measure a
metallicity for it, and no model spectrum is shown.

%%%%%%%%%%%%%%%%%%%%%%%%%%%%%%%%%
%%%%%%%%%   SECTION 4   %%%%%%%%%
%%%%%%%%%%%%%%%%%%%%%%%%%%%%%%%%%

\section{Spectroscopic Measurements}
\label{sec:spectroscopy}

\begin{deluxetable*}{lccccccr@{ }c@{ }lcccc}
\tablewidth{0pt}
\tablecolumns{12}
\tablecaption{Target List\label{tab:catalog}}
\tablehead{\colhead{ID} & \colhead{RA (J2000)} & \colhead{Dec (J2000)} & \colhead{$g_0$} & \colhead{$(g-r)_0$} & \colhead{$(g-I)_0$} & \colhead{S/N\tablenotemark{a}} & \multicolumn{3}{c}{$v_{\rm helio}$} & \colhead{Member?} & \colhead{$T_{\rm eff}$} & \colhead{$\log g$} & \colhead{[Fe/H]} \\
\colhead{ } & \colhead{ } & \colhead{ } & \colhead{(mag)} & \colhead{(mag)} & \colhead{(mag)} & \colhead{(\AA$^{-1}$)} & \multicolumn{3}{c}{(km~s$^{-1}$)} & \colhead{ } & \colhead{(K)} & \colhead{(cm~s$^{-2}$)} & \colhead{ }}
\startdata
\cutinhead{Hydra II}
196246                  & 12 21 27.50 & $-31$ 57 22.6 & 19.68 &  0.63 & \nodata &     \phn  54 & $ 161.8$ & $\pm$ & $ 1.6$ & N & \nodata & \nodata & \nodata \\
184647                  & 12 21 28.13 & $-32$ 06 36.7 & 19.75 &  0.63 & \nodata &     \phn  46 & $ 116.3$ & $\pm$ & $ 1.9$ & N & \nodata & \nodata & \nodata \\
199452                  & 12 21 30.18 & $-31$ 55 58.0 & 19.94 &  0.85 & \nodata &     \phn  59 & $   5.1$ & $\pm$ & $ 1.6$ & N & \nodata & \nodata & \nodata \\
189524                  & 12 21 32.70 & $-32$ 03 51.5 & 21.90 &  0.60 & \nodata & \phn\phn   8 & $ 272.9$ & $\pm$ & $ 7.9$ & N & \nodata & \nodata & \nodata \\
188838                  & 12 21 33.19 & $-32$ 04 29.4 & 20.66 &  0.67 & \nodata &     \phn  33 & $ 319.0$ & $\pm$ & $ 2.3$ & N & \nodata & \nodata & \nodata \\
194103                  & 12 21 34.74 & $-31$ 59 21.8 & 20.93 &  0.57 & \nodata &     \phn  25 & $ 300.0$ & $\pm$ & $ 4.2$ & Y & 5095 & 2.05 & $-2.40 \pm 0.32$ \\
193538                  & 12 21 35.93 & $-31$ 59 54.0 & 19.79 &  0.67 & \nodata &     \phn  66 & $ 304.2$ & $\pm$ & $ 1.7$ & Y & 4918 & 1.46 & $-1.95 \pm 0.12$ \\
198021                  & 12 21 36.31 & $-31$ 56 38.1 & 21.40 &  0.68 & \nodata &     \phn  19 & $  63.9$ & $\pm$ & $ 3.4$ & N & \nodata & \nodata & \nodata \\
195726                  & 12 21 36.74 & $-31$ 57 54.1 & 21.13 &  0.60 & \nodata &     \phn  22 & $ 306.5$ & $\pm$ & $ 3.5$ & Y & 5036 & 2.11 & $-2.48 \pm 0.33$ \\
190646                  & 12 21 38.18 & $-32$ 02 38.2 & 20.23 &  0.61 & \nodata &     \phn  44 & $  86.6$ & $\pm$ & $ 1.8$ & N & \nodata & \nodata & \nodata \\
194563                  & 12 21 38.51 & $-31$ 58 56.4 & 20.57 &  0.48 & \nodata &     \phn  34 & $ 302.9$ & $\pm$ & $ 3.3$ & Y & \nodata & \nodata & \nodata \\
191521                  & 12 21 39.68 & $-32$ 01 48.8 & 19.91 &  0.96 & \nodata &     \phn  45 & $ 202.5$ & $\pm$ & $ 1.8$ & N & \nodata & \nodata & \nodata \\
191385\tablenotemark{b} & 12 21 40.08 & $-32$ 01 57.7 & 20.70 &  0.16 & \nodata & \phn\phn   9 & $ 316.9$ & $\pm$ & $19.9$ & N & \nodata & \nodata & \nodata \\
192206                  & 12 21 40.37 & $-32$ 01 12.1 & 19.97 &  1.02 & \nodata &     \phn  68 & $  28.0$ & $\pm$ & $ 1.5$ & N & \nodata & \nodata & \nodata \\
193286                  & 12 21 40.96 & $-32$ 00 07.0 & 21.27 &  0.54 & \nodata &     \phn  14 & $ 299.9$ & $\pm$ & $ 5.4$ & Y & \nodata & \nodata & \nodata \\
194920                  & 12 21 41.02 & $-31$ 58 34.4 & 21.58 &  0.54 & \nodata &     \phn  16 & $ 294.5$ & $\pm$ & $ 5.9$ & Y & \nodata & \nodata & \nodata \\
194405                  & 12 21 41.05 & $-31$ 59 04.1 & 20.36 &  0.65 & \nodata &     \phn  48 & $ 304.2$ & $\pm$ & $ 1.9$ & Y & 4921 & 1.73 & $-1.89 \pm 0.13$ \\
189086                  & 12 21 41.79 & $-32$ 04 15.9 & 21.46 & -0.21 & \nodata & \phn\phn   4 & $ 315.5$ & $\pm$ & $23.5$ & Y & \nodata & \nodata & \nodata \\
194325                  & 12 21 42.51 & $-31$ 59 10.0 & 21.04 &  0.51 & \nodata &     \phn  24 & $ 301.9$ & $\pm$ & $ 3.0$ & Y & 5171 & 2.15 & $-2.76 \pm 0.43$ \\
197616                  & 12 21 42.62 & $-31$ 56 44.2 & 21.11 &  0.36 & \nodata &     \phn  15 & $  94.9$ & $\pm$ & $ 6.6$ & N & \nodata & \nodata & \nodata \\
201098                  & 12 21 42.78 & $-31$ 54 26.7 & 20.21 &  0.82 & \nodata &     \phn  22 & $ -85.0$ & $\pm$ & $ 2.3$ & N & \nodata & \nodata & \nodata \\
196797                  & 12 21 42.79 & $-31$ 57 04.2 & 22.37 &  0.43 & \nodata & \phn\phn   6 & $ 300.1$ & $\pm$ & $21.8$ & Y & \nodata & \nodata & \nodata \\
195247                  & 12 21 44.45 & $-31$ 58 21.0 & 21.92 &  0.48 & \nodata &     \phn  11 & $ 317.5$ & $\pm$ & $16.8$ & Y & \nodata & \nodata & \nodata \\
196052                  & 12 21 44.57 & $-31$ 57 37.6 & 21.85 &  0.56 & \nodata &     \phn  12 & $ 303.7$ & $\pm$ & $ 4.1$ & Y & \nodata & \nodata & \nodata \\
200162                  & 12 21 44.65 & $-31$ 55 17.7 & 19.34 &  0.98 & \nodata &     \phn  79 & $  68.1$ & $\pm$ & $ 1.5$ & N & \nodata & \nodata & \nodata \\
194736                  & 12 21 44.65 & $-31$ 58 47.6 & 21.95 &  0.49 & \nodata &     \phn  10 & $ 288.3$ & $\pm$ & $14.5$ & Y & \nodata & \nodata & \nodata \\
193869                  & 12 21 45.96 & $-31$ 59 34.4 & 20.11 &  0.50 & \nodata &     \phn  46 & $ 237.3$ & $\pm$ & $ 2.1$ & N & \nodata & \nodata & \nodata \\
183842                  & 12 21 46.10 & $-32$ 07 02.8 & 19.79 &  0.67 & \nodata &     \phn  55 & $ -19.6$ & $\pm$ & $ 1.7$ & N & \nodata & \nodata & \nodata \\
202029                  & 12 21 47.09 & $-31$ 53 36.0 & 21.69 &  0.50 & \nodata & \phn\phn   6 & $  10.1$ & $\pm$ & $ 9.0$ & N & \nodata & \nodata & \nodata \\
197129                  & 12 21 48.34 & $-31$ 56 52.2 & 21.50 &  0.62 & \nodata &     \phn  17 & $ 315.7$ & $\pm$ & $ 3.7$ & N & \nodata & \nodata & \nodata \\
192059                  & 12 21 48.39 & $-32$ 01 21.8 & 22.36 &  0.59 & \nodata & \phn\phn   7 & $ 172.4$ & $\pm$ & $10.0$ & N & \nodata & \nodata & \nodata \\
\cutinhead{Pisces II}
9004                    & 22 58 17.52 & $+05$ 55 17.5 & 20.29 &  0.73 & \nodata &     \phn  58 & $-224.9$ & $\pm$ & $ 1.6$ & Y & 4787 & 1.34 & $-2.38 \pm 0.13$ \\
9618                    & 22 58 20.35 & $+05$ 58 08.8 & 22.54 &  0.59 & \nodata & \phn\phn   8 & $-301.6$ & $\pm$ & $ 6.5$ & N & \nodata & \nodata & \nodata \\
9833                    & 22 58 21.22 & $+05$ 57 20.3 & 21.76 &  0.61 & \nodata &     \phn  19 & $-226.9$ & $\pm$ & $ 3.2$ & Y & \nodata & \nodata & \nodata \\
10215                   & 22 58 22.88 & $+05$ 57 35.5 & 21.53 &  0.65 & \nodata &     \phn  21 & $ -25.9$ & $\pm$ & $ 3.1$ & N & \nodata & \nodata & \nodata \\
10694                   & 22 58 25.06 & $+05$ 57 20.4 & 19.94 &  1.05 & \nodata &     \phn  59 & $-232.0$ & $\pm$ & $ 1.6$ & Y & 4132 & 0.81 & $-2.70 \pm 0.11$ \\
11592                   & 22 58 28.74 & $+05$ 56 56.9 & 19.91 &  1.08 & \nodata &     \phn  82 & $   4.3$ & $\pm$ & $ 1.5$ & N & \nodata & \nodata & \nodata \\
12924                   & 22 58 34.10 & $+05$ 57 43.6 & 21.19 &  0.63 & \nodata &     \phn  31 & $-221.6$ & $\pm$ & $ 2.9$ & Y & 4955 & 1.82 & $-2.10 \pm 0.18$ \\
13387                   & 22 58 35.91 & $+05$ 57 22.4 & 22.30 &  0.49 & \nodata &     \phn  11 & $-215.8$ & $\pm$ & $ 7.6$ & Y & \nodata & \nodata & \nodata \\
13560                   & 22 58 36.71 & $+05$ 58 06.2 & 21.48 &  0.58 & \nodata &     \phn  23 & $-232.6$ & $\pm$ & $ 5.3$ & Y & 5049 & 1.99 & $-2.15 \pm 0.28$ \\
13757                   & 22 58 37.49 & $+05$ 56 24.8 & 21.25 &  0.64 & \nodata &     \phn  27 & $-102.1$ & $\pm$ & $ 2.3$ & N & \nodata & \nodata & \nodata \\
14179                   & 22 58 39.36 & $+05$ 56 00.7 & 22.42 &  0.54 & \nodata & \phn\phn   8 & $-224.8$ & $\pm$ & $ 9.6$ & Y & \nodata & \nodata & \nodata \\
16716                   & 22 58 51.19 & $+05$ 59 57.1 & 20.33 &  0.74 & \nodata &     \phn  25 & $  -5.1$ & $\pm$ & $10.9$ & N & \nodata & \nodata & \nodata \\
17500                   & 22 58 54.78 & $+05$ 57 47.5 & 20.56 &  0.83 & \nodata &     \phn  25 & $  83.7$ & $\pm$ & $ 2.2$ & N & \nodata & \nodata & \nodata \\
\cutinhead{Laevens 1}
302                     & 11 36 13.55 & $-10$ 53 02.9 & 20.94 & \nodata &  1.29 &     \phn  20 & $ 149.8$ & $\pm$ & $ 2.9$ & Y & 4977 & 1.53 & $-1.59 \pm 0.15$ \\
374                     & 11 36 13.76 & $-10$ 52 48.5 & 19.90 & \nodata &  1.32 &     \phn  68 & $ 151.9$ & $\pm$ & $ 1.7$ & Y & 4966 & 1.09 & $-1.66 \pm 0.11$ \\
420                     & 11 36 13.90 & $-10$ 52 27.3 & 19.44 & \nodata &  1.59 &          105 & $ 149.8$ & $\pm$ & $ 1.5$ & Y & 4626 & 0.69 & $-1.55 \pm 0.11$ \\
378                     & 11 36 14.53 & $-10$ 52 43.7 & 21.55 & \nodata &  1.22 &     \phn  19 & $ 153.5$ & $\pm$ & $ 3.1$ & Y & \nodata & \nodata & \nodata \\
93                      & 11 36 15.91 & $-10$ 52 43.6 & 19.68 & \nodata &  1.52 &     \phn  88 & $ 147.2$ & $\pm$ & $ 1.6$ & Y & 4726 & 0.84 & $-1.65 \pm 0.11$ \\
1710                    & 11 36 16.52 & $-10$ 52 46.7 & 19.44 & \nodata &  1.31 &     \phn  16 & $ 143.6$ & $\pm$ & $ 4.5$ & Y & \nodata & \nodata & \nodata \\
1715                    & 11 36 16.59 & $-10$ 52 46.2 & 18.91 & \nodata &  0.85 &     \phn  66 & $  72.0$ & $\pm$ & $ 1.8$ & N & \nodata & \nodata & \nodata \\
367                     & 11 36 16.89 & $-10$ 52 43.7 & 22.54 & \nodata &  1.14 & \phn\phn   8 & $ 153.8$ & $\pm$ & $10.0$ & Y & \nodata & \nodata & \nodata \\
1717                    & 11 36 17.26 & $-10$ 52 46.2 & 18.99 & \nodata &  0.70 &     \phn  47 & $ 266.0$ & $\pm$ & $ 2.2$ & N & \nodata & \nodata & \nodata \\
1972                    & 11 36 17.70 & $-10$ 52 08.0 & 22.17 & \nodata &  1.04 &     \phn  10 & $ 141.4$ & $\pm$ & $ 4.7$ & Y & \nodata & \nodata & \nodata \\
1997                    & 11 36 18.37 & $-10$ 52 00.7 & 22.60 & \nodata &  1.07 & \phn\phn   6 & $ 166.6$ & $\pm$ & $11.5$ & N & \nodata & \nodata & \nodata \\
1684                    & 11 36 18.59 & $-10$ 52 49.7 & 21.23 & \nodata &  1.23 &     \phn  21 & $ 150.9$ & $\pm$ & $ 3.1$ & Y & 5064 & 1.70 & $-2.10 \pm 0.21$ \\
399                     & 11 36 19.58 & $-10$ 52 37.1 & 19.15 & \nodata &  0.70 &     \phn  80 & $ 155.3$ & $\pm$ & $ 1.8$ & N & \nodata & \nodata & \nodata \\
963                     & 11 36 20.11 & $-10$ 52 38.2 & 18.90 & \nodata &  1.87 &          150 & $ 149.2$ & $\pm$ & $ 1.5$ & Y & 4331 & 0.23 & $-1.78 \pm 0.11$ \\
\enddata
\tablenotetext{a}{To convert to S/N per pixel, multiply by 0.57.}
\tablenotetext{b}{Non-member based on CMD position.}
\end{deluxetable*}

\subsection{Radial Velocity Measurements}
\label{sec:rv}

We measured radial velocities of stars in the three Milky Way
satellites by comparing their spectra to the radial velocity template
stars' spectra.  In a manner similar to \citet{sim07}, we optimized
the DEEP2 survey's redshift measurement technique \citep{new13} for
measuring stellar velocities rather than galaxy redshifts.  This
technique involves computing $\chi^2 = ({\rm target~flux} - {\rm
  template~flux})^2 / ({\rm flux~error})^2$ for each template star.
The relative velocity between the target and the template is shifted
to find the minimum $\chi^2$.  The velocity of the star ($v_{\rm
  obs}$) is the one that minimizes $\chi^2$ among all of the
templates.  As \citeauthor{new13}\ pointed out, this technique is a
generalization of a cross-correlation \citep{ton79} that allows for
different values of flux error for each pixel.

Small astrometric errors or imprecision in the alignment of the
slitmask cause each star to fall at some displacement from the exact
center of the slit.  Mis-centering translates to a shift in the
wavelength scale.  Following \citet{sim07}, we corrected all of the
stellar velocities to a standard geocentric frame by finding the
velocity offset ($v_{\rm geo}$) required to align the observed
telluric absorption features with the features in the hot star
spectra.  The observed wavelength regions included in the velocity
correction were 6866--6912~\AA\ (B band), 7167--7320~\AA,
7593--7690~\AA\ (A band), and 8110--8320~\AA\@.  The velocity
corrections were determined with the $\chi^2$ technique described in
the previous paragraph.  The velocities quoted in this paper are in
the heliocentric frame: $v_{\rm helio} = v_{\rm obs} + v_{\rm geo} +
v_{\rm corr}$, where $v_{\rm corr}$ is the conversion from the
geocentric frame to the heliocentric frame.

We estimated random and systematic uncertainties separately.  To
estimate the random uncertainty, we resampled the spectra for $10^3$
Monte Carlo trials.  In one trial, the flux of each pixel was drawn
from a Gaussian probability distribution.  The mean of the
distribution was the original flux value, and its variance was the
flux variance estimated by \texttt{spec2d}.  We re-measured $v_{\rm
  helio}$ (including both $v_{\rm obs}$ and $v_{\rm geo}$) for all of
the trials, and we took the random uncertainty, $\delta_{\rm rand}v$,
to be the standard deviation of the $v_{\rm helio}$ measurements.

There is some minimum velocity uncertainty, even for noiseless
spectra.  This uncertainty can arise due to uncorrected flexure in
DEIMOS, spectral template mismatches, errors in the wavelength
solution, and unknown causes.  We estimated $\delta_{\rm sys}v$ by
calculating the difference in $v_{\rm helio}$ for two independent
measurements of the same set of stars.  The ideal sample for
estimating $\delta_{\rm sys}v$ will have many high-S/N spectra.
Instead of using the dwarf galaxy sample, we used spectra of \ngc\ red
giants in the GC M22\@.  The spectra were taken with the same slitmask
on consecutive nights (see Table~\ref{tab:obs}).  The data from each
night were reduced separately according to
Section~\ref{sec:reductions}.  We measured $v_{\rm helio}$ and
estimated $\delta_{\rm rand}v$ as described above.  We then calculated
the quantity $(v_{\rm helio,1}-v_{\rm helio,2})/\sqrt{\delta_{\rm
    rand,1}v^2 + \delta_{\rm rand,2}v^2 + 2\delta_{\rm sys}v^2}$ for
each pair of observations of the same star.  The standard deviation of
this quantity should be 1 for well-determined errors.  The value of
$\delta_{\rm sys}v$ required to satisfy that condition is
\vsyserr~km~s$^{-1}$.  The final error bar for each star is
$\sqrt{\delta_{\rm rand}v^2 + \delta_{\rm sys}v^2}$.

Our estimate of $\delta_{\rm sys}v$ ignores some potentially important
sources of error.  First, the repeat measurements were obtained with
the same slitmask.  As a result, some of the slit mis-centering---for
example, due to astrometric errors in the slitmask design---would be
repeated in the same way across the two consecutive nights.  Second,
the slitmasks were observed at about the same hour angle on each
night.  Although we applied wavelength corrections for flexure and
differential atmospheric refraction, these errors would have been of
the same sign and similar magnitude for both nights.  Any residual,
uncorrected error would be similar across both nights.  Third, the M22
spectra have fairly high S/N\@.  Consequently, the same template
spectrum was assigned to \ntemplate\ out of \ngc\ M22 stars.  This
fraction would be lower in spectra with lower S/N\@.

All of these effects would potentially lower the measurement of
$\delta_{\rm sys}v$ artificially.  In order to quantify the effect of
a spuriously low value of $\delta_{\rm sys}v$, we also calculated
velocity dispersions as described in Section~\ref{sec:sigmav} assuming
$\delta_{\rm sys}v = 2.2$~km~s$^{-1}$, the original value calculated
by \citet{sim07}, instead of \vsyserr~km~s$^{-1}$.  The increased
systematic error would make the upper limits on the velocity
dispersions about 5\% less stringent, and it would reduce the measured
velocity dispersion of Pisces~II by 13\%, well within the uncertainty.
In summary, our lower value of $\delta_{\rm sys}v$ than previous works
does not change our conclusions.

\begin{figure}[t!]
\centering
\includegraphics[width=\columnwidth]{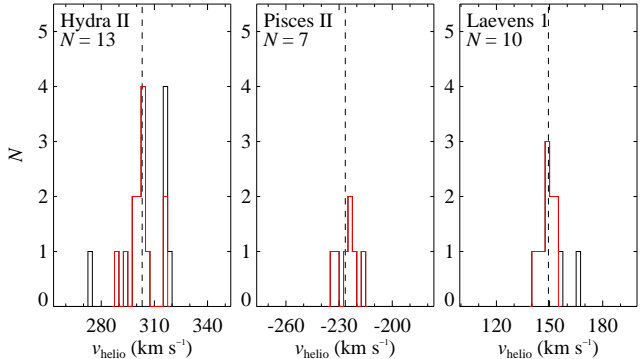}
\caption{The distributions of radial velocities for stars observed in
  the field of each satellite.  Stars in the red portion of the
  histograms are deemed members.  The dashed line shows $\langle
  v_{\rm helio} \rangle$, and $N$ is the number of member stars.  The
  membership cut includes 99\% of member stars ($2.58\sigma_v$ from
  the mean velocity).  Stars whose velocity error bars overlap any
  part of the allowed velocity range are considered
  members.\label{fig:v}}
\end{figure}

Table~\ref{tab:catalog} presents the velocities for stars in the
satellite galaxies.  The table excludes stars where velocity errors
exceeded 30~km~s$^{-1}$ and stars where velocity measurement was not
possible.  Figure~\ref{fig:v} shows velocity histograms for each
satellite in bins of 2.5~km~s$^{-1}$.  The domain of the plots is
100~km~s$^{-1}$ centered on the mean velocity for each satellite.  In
Section~\ref{sec:kinematics}, we use these velocities to measure the
kinematic properties of each satellite.

\subsection{Metallicity Measurements}
\label{sec:feh}

We measured metallicities by fitting synthetic spectra to the observed
spectra following the same procedure as \citet{kir08a,kir10}.  First,
we normalized each spectrum to its continuum.  To do so, we divided
the spectrum by a spline fit to spectral regions generally free of
absorption lines.  Then, we matched the spectrum to a grid of
synthetic spectra generated with ATLAS9 model atmospheres
\citep{kur93} and the synthesis code MOOG \citep{sne73}.  We measured
iron abundances, [Fe/H]\footnote{We adopted the solar abundances of
  \citet{and89} except for iron: $12 + \log (n({\rm Fe})/n({\rm H})) =
  7.52$ \citep{sne92}.}, by fitting only to iron absorption lines and
ignoring other spectral regions.  Please refer to \citet{kir10} for
more detail.

We estimated the random uncertainty on [Fe/H] from the covariance
matrix of the $\chi^2$ fit to the synthetic spectral grid.  There is
also a systematic uncertainty of 0.11~dex \citep{kir10}.  The final
error bar is the quadrature sum of the random and systematic
uncertainties.

\begin{figure}[t!]
\centering
\includegraphics[width=\columnwidth]{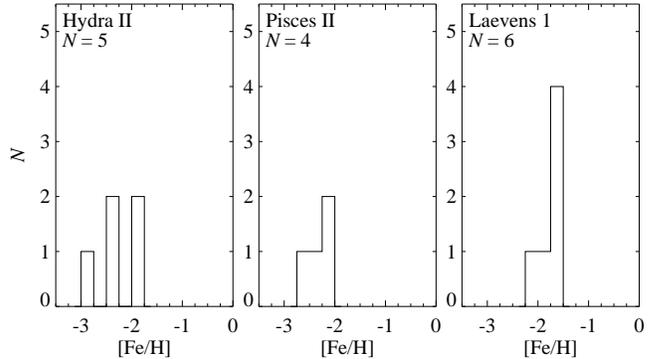}
\caption{The distribution of [Fe/H] in each satellite.  There are
  fewer stars in this figure than in Figure~\ref{fig:v} because we
  could not measure [Fe/H] for all radial velocity members.  $N$ is
  the number of stars for which [Fe/H] measurements with errors less
  than 0.5~dex were possible.\label{fig:feh}}
\end{figure}

Table~\ref{tab:catalog} gives effective temperatures ($T_{\rm eff}$),
surface gravities ($\log g$), and [Fe/H] measurements for member stars
with $\delta{\rm [Fe/H]} < 0.5$ and ${\rm S/N} > 20$~\AA$^{-1}$.
Figure~\ref{fig:feh} shows the distribution of these measurements in
bins of 0.25~dex.  By nature, ultra-faint satellites have only
handfuls of red giants bright enough for metallicity measurements.  As
a result, the metallicity distributions in Figure~\ref{fig:feh} have
only 4--6 stars.  Accurate quantification of metallicity distributions
and chemical evolution requires larger samples.  Nonetheless, we
present rudimentary metallicity averages and dispersions in
Section~\ref{sec:metallicity}.

\subsection{Comparison to Other Measurements}
\label{sec:comparison}

\begin{figure}[t!]
\centering
\includegraphics[width=\columnwidth]{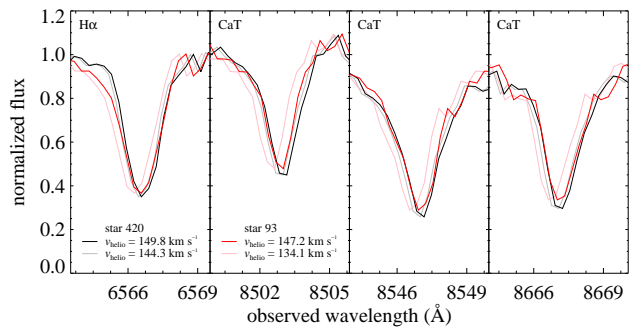}
\caption{DEIMOS spectra of Laevens~1 stars 420 (black) and 93 (red),
  which were also observed by \citet{bonif15}.  The bold (black and
  red) spectra show the spectra as we observed them.  The faded (gray
  and pink) spectra show the spectra as they would have appeared at
  the radial velocities measured by \citeauthor{bonif15} The visible
  separation between the faded spectra shows that our spectra have the
  precision to rule out a velocity separation of
  10.2~km~s$^{-1}$.\label{fig:comparison}}
\end{figure}

\citet{bonif15} obtained VLT/X-Shooter spectra of two stars in
Laevens~1.  Both stars are also in our spectroscopic sample.  We call
them 420 and 93, and \citeauthor{bonif15}\ call them J113613$-$105227
and J113615$-$105244.  They measured a velocity difference between the
two stars of $10.2 \pm 5.7$~km~s$^{-1}$, whereas we measured a
difference of only $\lvdiff \pm \lvdifferr$~km~s$^{-1}$.

Figure~\ref{fig:comparison} shows our DEIMOS spectra in the observer
frame.  The bold black and red spectra show the spectra as we observed
them, and the faded (gray and pink) lines show the spectra as they
would have appeared at the velocities measured by \citet{bonif15}.
The bold spectra are nearly superposed on top of each other, whereas
the faded spectra are visibly separated.  Thus, the figure shows that
our spectra have the precision to discern between a 10.2~km~s$^{-1}$
and a \lvdiff~km~s$^{-1}$ difference.

There are several possible reasons for the discrepancy between our two
studies.  First, one or both of the stars could have a variable radial
velocity.  In that case, we happened to observe the two stars when
they had nearly the same radial velocity, and our inability to resolve
the velocity dispersion (Section~\ref{sec:sigmav}) is unaffected.
Second, one of our studies could be subject to a systematic error.
For example, we corrected for slit mis-centering by applying a
velocity shift based on telluric absorption.  (The wavelength scale in
Figure~\ref{fig:comparison} includes the telluric correction.)
\citeauthor{bonif15}\ were unable to apply a similar correction.
Therefore, one possible systematic error in their study was slightly
different positions of the two stars perpendicular to the X-Shooter
slit.  Regardless, \citeauthor{bonif15}\ state that the velocity
dispersion of Laevens~1 is consistent with zero when the measurement
errors are taken into account.  This basic result agrees with our own
conclusions (Section~\ref{sec:sigmav}).

We measured ${\rm [Fe/H]} = \lonefeh \pm \lonefeherr$ and $\ltwofeh
\pm \ltwofeherr$ for stars 420 and 93, respectively.
\citeauthor{bonif15}\ measured ${\rm [Fe/H]} = -1.73 \pm 0.26$ and
$-1.67 \pm 0.28$ from an equivalent width analysis using MyGIsFOS
\citep{sbo14}.  Our measurements are entirely consistent within the
measurement uncertainty.

%%%%%%%%%%%%%%%%%%%%%%%%%%%%%%%%%
%%%%%%%%%   SECTION 5   %%%%%%%%%
%%%%%%%%%%%%%%%%%%%%%%%%%%%%%%%%%

\section{Kinematics}
\label{sec:kinematics}

\subsection{Membership}
\label{sec:membership}

Ultra-faint satellites have very low surface brightness.  In some
cases, most of the point sources in the vicinity of an ultra-faint
satellite---even at the center---are foreground dwarf stars.  We
avoided most of the contamination by selecting stars with colors and
magnitudes appropriate for red giants at the distance of each
satellite (Section~\ref{sec:phot} and Figure~\ref{fig:cmds}).  We
removed the remainder of the contaminants by imposing a radial
velocity cut.

We determined membership in each satellite by using its mean velocity,
$\langle v_{\rm helio} \rangle$, and velocity dispersion, $\sigma_v$.
First, we started with guesses for $\langle v_{\rm helio} \rangle$ and
$\sigma_v$.  We required member stars to have $|v_{\rm helio} -
\langle v_{\rm helio} \rangle| < 2.58\sigma_v$.  For a Gaussian
velocity distribution, this cut includes 99\% of member stars.
Because $\sigma_v$ for most of the satellites is very small, a hard
cut would exclude low-S/N stars whose measured velocities are
discrepant by more than $2.58\sigma_v$ simply because their
uncertainties are larger than that.  Therefore, we extend the
membership criterion to any star whose $1\sigma$ error bar overlaps
the velocity range for member stars.  We determined $\langle v_{\rm
  helio} \rangle$ and $\sigma_v$ from the list of member stars
following the procedure in Section~\ref{sec:sigmav}.  Then, we used
these new values to reform the member list.  We repeated this
procedure until the member list did not change from one iteration to
the next.

Foreground dwarfs can also be identified by spectral features
sensitive to surface gravity, such as the Na~{\sc i} doublet at
8190~\AA\ \citep{spi71,cohen78} and Mg~{\sc i}~8807 \citep{bat12}.  We
found several stars with very strong Na doublets and Mg~{\sc i} lines
among the Hydra~II and Pisces~II samples, but they were already
excluded by the radial velocity cut.  Hence, we did not need to impose
any additional membership criteria.  We discuss Mg~{\sc i} again in
Section~\ref{sec:sigmav} in reference to a star that barely missed the
membership cut in Hydra~II\@.

Two of the three blue stars that we targeted in Laevens~1 are
non-members on the basis of their radial velocities.  \citet{bel14}
speculated that these stars might be blue loop stars.  If they were,
then Laevens~1 must have a young stellar population, which would argue
strongly that it is a star-forming galaxy, not a GC\@.  However, the
non-membership of these two stars negates most of the evidence that
Laevens~1 has a very young stellar population, although
\citet{bonif15} argued that the blue extension of the main sequence
might indicate the presence of a $\sim$2~Gyr old population.  The
third blue star, 399, has $v_{\rm helio} = \vcrtblue \pm
\vcrtblueerr$~km~s$^{-1}$, which is $\vcrtbluediff$~km~s$^{-1}$ from
the mean radial velocity.  This star formally passes the velocity
membership cut.  However, in addition to its unusual color, it has a
more discrepant radial velocity than any of the other member stars,
and it is farther from the center of the cluster than all but one
confirmed member.  We considered the possibility that this star is a
Cepheid member of Laevens~1.  However, the Pan-STARRS1 survey
\citep{kai10} obtained six observations of this star in each of the
$r_{\rm P1}$, $z_{\rm P1}$, and $y_{\rm P1}$ bands, and the rms
scatter of these observations is consistent with measurement noise
($\sim0.03$ mag, B.~Sesar, priv.~comm.).  Since Population~II Cepheids
have light curve amplitudes between 0.5 and 1.2~mag in the Johnson
$R-$band \citep{wal84}, we conclude that the blue star is not a
Cepheid variable.  We discuss the impact that including star 399 as a
member would have on $\sigma_v$ in Section~\ref{sec:sigmav}.

\subsection{Mean Velocities, Velocity Dispersions, and Masses}
\label{sec:sigmav}

We estimated $\langle v_{\rm helio} \rangle$ and $\sigma_v$ for the
three satellites in the same manner that \citet{kir13a} measured these
values for the UFD Segue~2.  That method, in turn, was based on
\citeauthor{wal06}'s (\citeyear{wal06}) procedure for measuring the
velocity dispersion of the Fornax dwarf spheroidal galaxy (dSph).  The
method uses maximum likelihood statistics and a Monte Carlo Markov
chain (MCMC)\@.  We maximized the logarithm of the likelihood ($L$)
that the given values of $\langle v_{\rm helio} \rangle$ and
$\sigma_v$ described the observed velocity distribution, including the
uncertainty estimates for individual stars.

\begin{eqnarray}
\nonumber \log L &=& \frac{N \log(2 \pi)}{2} + \frac{1}{2} \sum_i^N \left(\log((\delta v_{\rm helio})_i^2 + \sigma_v^2\right) \\
& & + \frac{1}{2} \sum_i^N \left(\frac{((v_{\rm helio})_i - \langle v_{\rm helio} \rangle)^2}{(\delta v_{\rm helio})_i^2 + \sigma_v^2}\right) \label{eq:v}
\end{eqnarray}

\noindent As initial guesses for $\langle v_{\rm helio} \rangle$ and
$\sigma_v$, we started with the mean and standard deviation of $v_{\rm
  helio}$ over the velocity range shown in Figure~\ref{fig:v}.  The
final results are not sensitive to these guesses.  We explored the
parameter space with a Metropolis-Hastings implementation of an
MCMC\@.  The length of the chain was $10^7$ trials.

As a test of our procedure, we measured $\langle v_{\rm helio} \rangle
= \gcmeanv \pm \gcmeanverr$~km~s$^{-1}$ and $\sigma_v =
\gcsigmav_{-\gcsigmaverrlower}^{+ \gcsigmaverrupper}$~km~s$^{-1}$ for
M22 from the 2009 Oct 14 observation.  In comparison, \citet{pet94}
measured $\langle v_{\rm helio} \rangle = -148.8 \pm 0.8$~km~s$^{-1}$
and $\sigma_v = 6.6 \pm 0.8$~km~s$^{-1}$ within $7\arcmin$ of the
cluster center, and \citet{lan09} measured $\langle v_{\rm helio}
\rangle = -144.9 \pm 0.3$~km~s$^{-1}$ and a central dispersion of
$\sigma_v = 6.8 \pm 0.9$~km~s$^{-1}$.  Our measurement of $\langle
v_{\rm helio} \rangle$ is consistent with that of \citeauthor{lan09},
and our measurement of $\sigma_v$ is consistent with both
\citeauthor{pet94}\ and \citeauthor{lan09} The velocity dispersion in
M22 decreases to about 5~km~s$^{-1}$ at $8\arcmin$ from the cluster
center \citep{lan09}.  However, our data is largely insensitive to the
decline is $\sigma_v$ because the maximum radial extent of our
spectroscopy is $8\arcmin$, and most of our M22 targets are within
$5\arcmin$.

Table~\ref{tab:properties} gives $\langle v_{\rm helio} \rangle$ and
$\sigma_v$ with errors.  It also gives velocities relative to the
Galactic standard of rest (GSR) assuming that the Sun's orbital
velocity is 220~km~s$^{-1}$.  Hydra~II is receding from the Galactic
center.  Therefore, it is past its pericenter on the way to its
apocenter.  Its heliocentric radial velocity is also similar to that
of the gas in the leading arm of the Magellanic stream
\citep{put98,bru05,nid08}.  This finding strengthens the potential for
Hydra~II to be associated with the LMC \citep{kop15a,bec15,dea15}.
Pisces~II is approaching the Galactic center, which means that it is
on the way to its pericenter.  Finally, Laevens~1 has a very small
$v_{\rm GSR}$, meaning that it is at pericenter or apocenter.  Because
it is so distant (at least 140~kpc from the Galactic center), it is
either close to apocenter or on a quasi-circular orbit.

\begin{figure}[t!]
\centering
\includegraphics[width=\columnwidth]{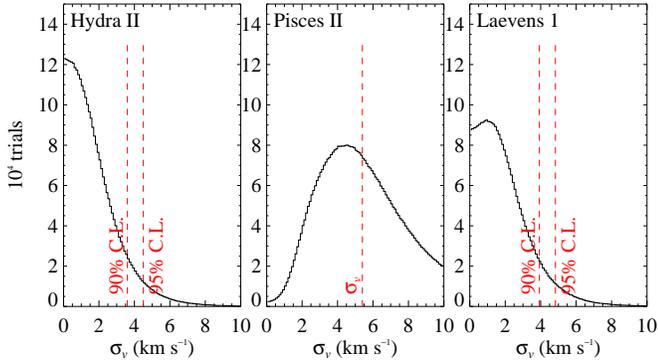}
\caption{The probability distributions for $\sigma_v$.  The histograms
  show successful MCMC trials.  Vertical lines show the measured
  velocity dispersion (Pisces~II) or upper limits (Hydra~II and
  Laevens~1).\label{fig:sigmav_confidence}}
\end{figure}

\begin{deluxetable*}{lccc}
\tablewidth{0pt}
\tablecolumns{4}
\tablecaption{Satellite Properties\label{tab:properties}}
\tablehead{\colhead{Property} & \colhead{Hydra~II} & \colhead{Pisces~II} & \colhead{Laevens 1}}
\startdata
$N_{\rm member}$ & 13 &  7 & 10 \\
$\log (L_{V}/L_{\sun})$ & $3.90 \pm 0.10$ & $3.93 \pm 0.20$ & $3.65 \pm 0.08$ \\
$r_h$ (arcmin) & $1.7_{-0.2}^{+0.3}$ & $1.1 \pm 0.1$ & $0.47_{-0.03}^{+0.04}$ \\
$r_h$ (pc) & $ 66_{-  9}^{+ 12}$ & $ 58 \pm   7$ & $ 19 \pm   2$ \\
$\langle v_{\rm helio} \rangle$ (km~s$^{-1}$) & $ 303.1 \pm 1.4$ & $-226.5 \pm 2.7$ & $ 149.3 \pm 1.2$ \\
$v_{\rm GSR}$ (km~s$^{-1}$) & $ 135.4$ & $ -79.9$ & $   4.6$ \\
$\sigma_v$ (km~s$^{-1}$) & $<3.6$ (90\% C.L.) & $5.4_{-2.4}^{+3.6}$ & $<3.9$ (90\% C.L.) \\
 & $<4.5$ (95\% C.L.) & & $<4.8$ (95\% C.L.) \\
$\log (M_{1/2}/M_{\sun})$ & $<5.9$ (90\% C.L.) & $6.2_{-0.2}^{+0.3}$ & $<5.5$ (90\% C.L.) \\
 & $<6.1$ (95\% C.L.) & & $<5.6$ (95\% C.L.) \\
$(M/L_V)_{1/2}$\tablenotemark{a} ($M_{\sun}/L_{\sun}$) & $< 200$ (90\% C.L.) & $ 370_{-240}^{+310}$ & $< 130$ (90\% C.L.) \\
 & $< 315$ (95\% C.L.) & & $< 192$ (95\% C.L.) \\
$\langle {\rm [Fe/H]} \rangle$ & $-2.02 \pm 0.08$ & $-2.45 \pm 0.07$ & $-1.68 \pm 0.05$ \\
$\sigma({\rm [Fe/H]})$ & $0.40_{-0.26}^{+0.48}$ & $0.48_{-0.29}^{+0.70}$ & $<0.40$ (90\% C.L.) \\
 & & & $<0.53$ (95\% C.L.) \\
\enddata
\tablenotetext{a}{Mass-to-light ratio within the half-light radius.}
\tablerefs{The measurements of $\log L_V$ and $r_h$ come from \citet{mar15}, \citet{bel10}, and \citet{lae14}.}
\end{deluxetable*}

Figure~\ref{fig:sigmav_confidence} shows the distribution of
$\sigma_v$ for accepted MCMC trials.  These distributions are
equivalent to the probability distribution for $\sigma_v$.  The
distribution for Pisces~II has a well-defined peak separated from
zero.  The dashed line shows the median value of $\sigma_v =
\sigmavp$~km~s$^{-1}$.  We estimated asymmetric error bars by
calculating the values of $\sigma_v$ that bracket 68.3\% of the
probability distribution.

The distributions for Hydra~II and Laevens~1 are concentrated near
zero.  Hence, our measurements cannot resolve $\sigma_v$ for these two
satellites.  We estimated two sets of upper limits by finding the
value of $\sigma_v$ that exceeds 90\% and 95\% of the MCMC trials.
Table~\ref{tab:properties} gives both upper limits for both
satellites.

Hydra~II and Laevens~1 have three stars with somewhat ambiguous
membership.  Stars 197129 and 188838 in Hydra~II have velocities
larger than $\langle v_{\rm helio} \rangle$ by $\vnmhtwodiff$ and
$\vnmhonediff$~km~s$^{-1}$.  If our membership cut were more inclusive
than $2.58\sigma_v$, then these stars could be considered members.
Unfortunately, neither star has a visible Na~{\sc i} doublet to help
define its status.  The Mg~{\sc i} equivalent width (EW) of star
188838 is $0.34 \pm 0.07$~\AA, and the combined EW of of the two
redder lines of the Ca~{\sc ii} infrared triplet (CaT) is $4.1 \pm
0.3$~\AA\@.  These measurements fall right on the dividing line
between dwarfs and giants defined by \citet{bat12}.  Because most
metal-poor dSph stars fall well below the dividing line, this test
disfavors membership for star 188838.  The Mg~{\sc i} line in star
197129 is too noisy to be useful.

Including star 192179 as a member does not change our qualitative
conclusions.  Instead, the 90\% C.L.\ upper limit on $\sigma_v$ rises
from $\sigmavlimoneh$~km~s$^{-1}$ to $\sigmavlimmodtwoh$~km~s$^{-1}$.
On the other hand, including star 188838 resolves the velocity
dispersion as $\sigma_v =
\sigmavmodoneh_{-\sigmaverrlowermodoneh}^{+\sigmaverruppermodoneh}$~km~s$^{-1}$.
Including both stars results in $\sigma_v =
\sigmavmodthreeh_{-\sigmaverrlowermodthreeh}^{+\sigmaverruppermodthreeh}$~km~s$^{-1}$.
However, both stars would be $>2\sigma_v$ outliers even with the
larger $\sigma_v$.  Furthermore, Figure~\ref{fig:cmds} shows that both
stars are redder relative to the M92 isochrone than any other member.
Figure~\ref{fig:maps} also shows that star 188838 would be the most
distant member of Hydra~II\@.  Because the majority of the evidence
disfavors membership, we ruled both stars as non-members.  Including
or excluding the two possible HB stars, 189086 and 191385, changes the
limits on $\sigma_v$ by less than 2\%.  Based on their CMD positions,
we ruled 189086 a member and 191385 a non-member.

Including star 399 in Laevens~1 resolves the velocity dispersion as
$\sigma_v =
\sigmavmodc_{-\sigmaverrlowermodc}^{+\sigmaverruppermodc}$~km~s$^{-1}$.
As discussed in Section~\ref{sec:membership}, the star is bright and
blue.  As a result, star 399 is probably not a member.

The total mass of a spherical stellar system in dynamical equilibrium
is related to the square of the velocity dispersion.  \citet{wol10}
showed that the total mass is poorly constrained because any possible
underlying dark matter has an unknown mass profile, and the velocity
anisotropy cannot be determined from a small sample of radial
velocities.  However, the mass within the half-light radius,
$M_{1/2}$, is well-constrained: $M_{1/2} = 4G^{-1}\sigma_v^2r_h$.
Although $r_h$ is the 2-D half-light radius, the formula infers the
mass enclosed within the 3-D half-light radius.

Table~\ref{tab:properties} gives $\log M_{1/2}$ for Pisces~II and
upper limits for Hydra~II and Laevens~1, which have only upper limits
for $\sigma_v$.  Pisces~II has a dynamical mass on par with dwarf
galaxies of similar luminosity \citep{str08,wol10}.  For example, its
luminosity and mass are very similar to Canes Venatici~II
\citep{sim07}.  The mass limits for Hydra~II and Laevens~1 are not
stringent enough to make these satellites unusually light.  The 90\%
C.L.\ for Hydra~II is $M_{1/2} < 1.0 \times 10^6~M_{\sun}$, which is
less than a factor of 2 smaller than Pisces~II\@.  The 90\% C.L.\ mass
limit for Segue~2, the least massive galaxy, is 7 times more stringent
than Hydra~II\@.  The limit for Laevens~1 is slightly more stringent
than Hydra~II, but we argue below that Laevens~1 is not a galaxy.

\citet{wil12} proposed that a ``galaxy'' be defined as a stellar
system that cannot be explained by a combination of baryons and
Newton's laws of gravity.  The mass-to-light ratio for Pisces~II is
$\mlp_{-\mlerrlowerp}^{+\mlerrupperp}~M_{\sun}/L_{\sun}$.  This value
is far too large to be explained by baryons alone, even for the oldest
stellar populations in the Universe.  In fact, $\mlruleoutp\%$ of the
successful MCMC trials have $M/L > 10~M_{\sun}/L_{\sun}$.  Hence,
Pisces~II satisfies the definition of a galaxy.  However, upper limits
on the mass-to-light ratios on Hydra~II and Laevens~1 do not help in
deciding if they are galaxies.  In Section~\ref{sec:metallicity}, we
use chemical evidence to resolve the nature of these two satellites.

One of the reasons dwarf galaxies are interesting is that they are
targets for the detection of gamma rays due to dark matter
self-annihilation.  Pisces~II has similar structural properties ($r_h$
and $\sigma_v$) and distance to Canes Venatici~II\@.  Hence, the two
galaxies' potential for the detection of the gamma ray signal is about
the same.  \citet{bon15} found that Canes Venatici~II is not the most
promising dwarf galaxy to search for self-annihilation, but it would
contribute significantly to an analysis that stacks the {\it Fermi}
gamma ray telescope observations of all of the dwarfs
\citep[e.g.,]{ack14,fermi15}.

%%%%%%%%%%%%%%%%%%%%%%%%%%%%%%%%%
%%%%%%%%%   SECTION 6   %%%%%%%%%
%%%%%%%%%%%%%%%%%%%%%%%%%%%%%%%%%

\section{Metallicity}
\label{sec:metallicity}

The stellar mass of a UFD is insufficient to retain supernova ejecta.
Nonetheless, all UFDs studied in sufficient detail show evidence for
metallicity dispersions \citep[e.g.,][]{fre10b,nor10,kir13a,var13}.
Their ability to self-enrich with iron implies that they have or once
had enough mass to prevent metal-enriched supernova ejecta from
escaping.  As a result, \citet{wil12} considered a dispersion in
[Fe/H] sufficient for classification as a galaxy for most stellar
systems.\footnote{Some massive GCs, like $\omega$~Centauri
  \citep{nor95}, have a dispersion in metallicity but no kinematic
  evidence for dark matter.}

We calculated the metallicity mean and dispersion in the same way that
we computed $\langle v_{\rm helio} \rangle$ and $\sigma_v$.  In
analogy to Equation~\ref{eq:v}, we maximized the likelihood that the
metallicity distribution has a mean $\langle {\rm [Fe/H]} \rangle$ and
dispersion $\sigma({\rm [Fe/H]})$ as follows:

\begin{eqnarray}
\nonumber \log L &=& \frac{N \log(2 \pi)}{2} + \frac{1}{2} \sum_i^N \left(\log(\delta {\rm [Fe/H]})_i^2 + \sigma({\rm [Fe/H]})^2\right) \\
& & + \frac{1}{2} \sum_i^N \left(\frac{({\rm [Fe/H]}_i - \langle {\rm [Fe/H]} \rangle)^2}{(\delta {\rm [Fe/H]})_i^2 + \sigma({\rm [Fe/H]})^2}\right) \label{eq:feh}
\end{eqnarray}

\noindent We used the metallicity measurements shown in
Table~\ref{tab:catalog}, i.e., those with $\delta {\rm [Fe/H]} < 0.5$
and ${\rm S/N} > 20$~\AA$^{-1}$.  The values were determined through
$10^7$ MCMC trials.

\begin{figure}[t!]
\centering
\includegraphics[width=\columnwidth]{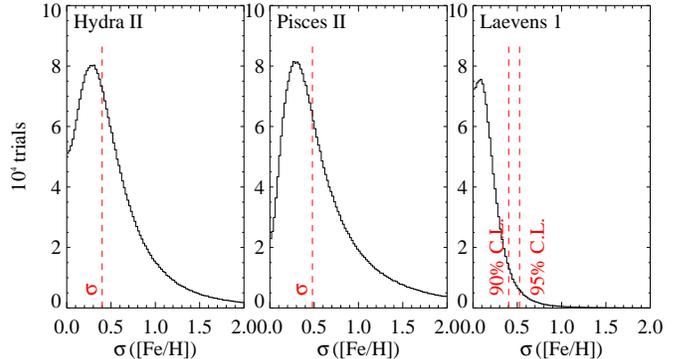}
\caption{The probability distributions for $\sigma({\rm [Fe/H]})$\@.
  The histograms show successful MCMC trials.  Vertical lines show the
  measured metallicity dispersions (Hydra~II and Pisces~II) or upper
  limits (Laevens~1).\label{fig:sigmafeh_confidence}}
\end{figure}

Figure~\ref{fig:sigmafeh_confidence} shows the distribution of
$\sigma({\rm [Fe/H]})$ for the successful MCMC trials.  Whereas the
distributions for Hydra~II and Pisces~II are somewhat separated from
zero, the distribution for Laevens~1 piles up at zero.  Hence, we
marginally resolve the metallicity dispersion for Hydra~II and
Pisces~II: \sigmafehruleouth\% and \sigmafehruleoutp\% of the MCMC
trials have $\sigma_{\rm [Fe/H]} > 0.2$, respectively.  On the other
hand, we measured only an upper limit for Laevens~1\@.  These values
are shown in Table~\ref{tab:properties}.

Although we did not resolve the velocity dispersion for Hydra~II, we
did detect a metallicity dispersion, albeit with somewhat low
significance.  Because it seems that Hydra~II enriched itself with
iron, we tentatively assign it a classification as a galaxy.  This
spectroscopic classification supports the photometric classification
based on its large half-light radius \citep{mar15}.  Pisces~II also
has a marginally resolved metallicity dispersion, consistent with its
dynamical classification as a galaxy.

\begin{figure}[t!]
\centering
\includegraphics[width=\columnwidth]{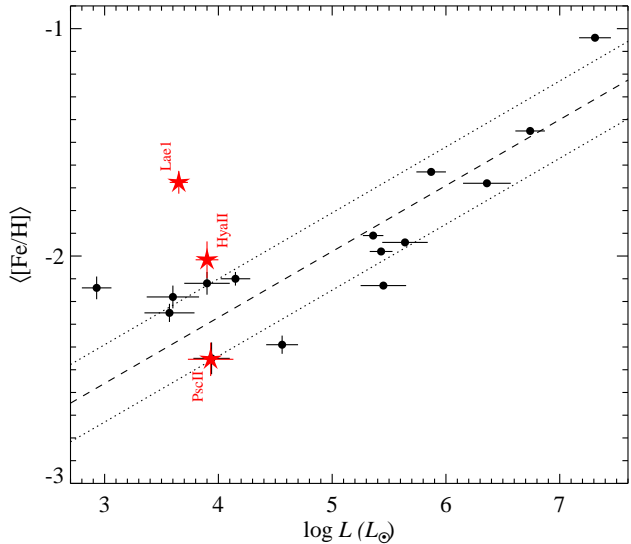}
\caption{The luminosity--metallicity relation for Milky Way satellite
  galaxies.  Hydra~II and Pisces~II fall in the midst of other UFDs.
  Pisces~II falls almost directly on top of Leo~IV\@.  Like other GCs,
  Laevens~1 does not conform to the relation.  The black points are
  $\langle {\rm [Fe/H]} \rangle$ measurements for Milky Way dSphs from
  \citet{kir13b}.  The dashed line is a least-squares fit to the black
  points excluding Segue~2, the least luminous galaxy on the plot.
  The dotted lines show the rms dispersion about the
  relation.\label{fig:lzr}}
\end{figure}

Dwarf galaxies obey a tight relationship between luminosity and
average metallicity \citep{ski89,mat98,kir11,kir13b}.  In fact, the
luminosity--metallicity relation (LZR) can be used as a diagnostic for
whether a stellar system is a galaxy.  Figure~\ref{fig:lzr} shows the
LZR from \citet{kir13b} with the three new satellites.  Hydra~II and
Pisces~II lie in the same region of the diagram as other UFDs, like
Leo~IV and Canes Venatici~II\@.  Thus, the LZR supports their
identification as galaxies.

On the other hand, Laevens~1 is too metal-rich for its luminosity (or
too faint for its metallicity).  GCs do not obey any LZR\@.  Tidal
dwarf galaxies also do not obey the LZR, but tidal dwarfs tend to be
close to solar metallicity \citep{duc00}.  Therefore, Laevens~1's
position in Figure~\ref{fig:lzr} suggests that it is a GC or a
severely tidally stripped dwarf galaxy, in which the removal of stars
has reduced Laevens~1's mass but not its metallicity.  \citet{kir13a}
proposed this scenario for Segue~2, but Segue~2 has a metallicity
dispersion.  We detected no metallicity dispersion in Laevens~1,
favoring its identification as a GC\@.

Laevens~1 has four pieces of evidence that suggest that it is a GC,
not a galaxy.  First, it has a smaller half-light radius than known
UFDs \citep{lae14}.  Second, it does not have strong evidence for dark
matter ($M/L < \mllimonec~M_{\sun}/L_{\sun}$, 90\% C.L.)\@.  Third,
its metallicity dispersion is less than \sigmafehlimonec~dex (90\%
C.L.)\@.  Fourth, it does not obey the LZR for dwarf galaxies.
Additionally, most of the blue stars that may have suggested that
Laevens~1 has a young stellar population turned out to be probable
non-members (Section~\ref{sec:membership}).  The various lines of
evidence taken together favor a cluster classification rather than a
dwarf galaxy.

Star~10694, the brightest member of Pisces~II, is carbon-rich.
Figure~\ref{fig:spectra} shows CN absorption between 8300~\AA\ and
8400~\AA\@.  Bright red giants should be destroying carbon, not
creating it.  Star~10694 could be an AGB star that is currently
dredging up carbon.  Alternatively, it could have acquired carbon from
a recently defunct AGB companion.  Its large luminosity favors the
scenario that it is itself an AGB star.

%%%%%%%%%%%%%%%%%%%%%%%%%%%%%%%%%
%%%%%%%%%   SECTION 7   %%%%%%%%%
%%%%%%%%%%%%%%%%%%%%%%%%%%%%%%%%%

\section{Summary}
\label{sec:summary}

We obtained Keck/DEIMOS spectroscopy for three newly discovered Milky
Way satellites: Hydra~II, Pisces~II, and Laevens~1\@.  We measured
radial velocities and metallicities for individual candidate member
stars.  We identified \nh, \np, and \nc\ member stars in the three
satellites, respectively, on the basis of their radial velocities.
Most of the members are red giants, although there is one HB and one
AGB star in Hydra~II and one carbon-rich star in Pisces~II\@.

We could not resolve the velocity dispersion of Hydra~II, but we did
measure a non-zero dispersion in metallicity ($\sigma({\rm [Fe/H]}) =
\fehsigmah_{-\fehsigmaerrlowerh}^{+\fehsigmaerrupperh}$).  Because it
seems to have chemically enriched itself in iron and because it has a
large half-light radius \citep[66~pc,][]{bel10}, Hydra~II is more
likely to be a dwarf galaxy than a GC\@.  It also has a radial
velocity similar to the gas in the leading arm of the Magellanic
stream at its location.  Therefore, it may have fallen into the Milky
Way with the Magellanic Clouds.

Pisces~II is a bona fide galaxy that inhabits a massive dark matter
halo.  We measured a velocity dispersion in Pisces~II that is far in
excess of what would be expected on the basis of its stellar mass
alone.  We also marginally resolved a metallicity dispersion
($\sigma({\rm [Fe/H]}) =
\fehsigmap_{-\fehsigmaerrlowerp}^{+\fehsigmaerrupperp}$), indicating
that Pisces~II is not only massive now, but it was also massive enough
during star formation to retain supernova ejecta.

Laevens~1 is more likely a GC than a galaxy.  We did not resolve a
dispersion in velocity or metallicity, and Laevens~1 does not obey the
LZR for dwarf galaxies, whereas Hydra~II and Pisces~II do.  Although
the upper limit on the mass-to-light ratio ($M/L <
\mllimonec~M_{\sun}/L_{\sun}$, 90\% C.L.) does not rule out that
Laevens~1 could be dark matter-dominated, the low metallicity
dispersion and the inconsistency with the LZR suggest that the system
is a globular cluster.

\citet{bel14} found several bright, blue stars in the vicinity of
Laevens~1\@.  They suggested that these could be blue loop stars,
which would signify the presence of a very young stellar population in
Laevens~1\@.  However, we found these stars to be probable non-members
on the basis of radial velocity.

With this work, we have spectroscopically confirmed two dwarf galaxies
and one GC\@.  The dwarf galaxies are especially interesting for their
application to dark matter physics.  In particular, Pisces~II has a
mass-to-light ratio within the half-light radius of
$\mlp_{-\mlerrlowerp}^{+\mlerrupperp}~M_{\sun}/L_{\sun}$.  Although
this does not make it one of the most dark matter-dominated galaxies
known, it is still massive enough to warrant inclusion in a search for
gamma rays due to dark matter self-annihilation
\citep[e.g.,][]{ack14,fermi15,bon15}.

\acknowledgments We thank D.~Perley for obtaining LRIS images of
Laevens~1, J.~A.~Newman for assistance with the DEIMOS wavelength
solution, E.~Tollerud for the spectrum of the radial velocity template
star HD~38230, and M.~de los Reyes for pointing out that tidal dwarf
galaxies do not obey the LZR\@.  We also thank B.~Sesar and the
Pan-STARRS team for information regarding star 399 in Laevens~1.  The
referee's helpful comments significantly improved this article.

We are grateful to the many people who have worked to make the Keck
Telescope and its instruments a reality and to operate and maintain
the Keck Observatory.  The authors wish to extend special thanks to
those of Hawaiian ancestry on whose sacred mountain we are privileged
to be guests.  Without their generous hospitality, none of the
observations presented herein would have been possible.

{\it Facility:} \facility{Keck:I (LRIS), Keck:II (DEIMOS)}

\bibliography{dwarfs2015}
\bibliographystyle{apj}

\end{document}